\documentclass[aps,prl,preprint,groupedaddress]{revtex4}
\usepackage{graphicx}
\begin{document}

\title{Prediction of Half Metallicity along the Edge of Boron Nitride Zigzag Nanoribbons}
\author{Fawei Zheng$^1$, Gang Zhou$^1$, Zhirong Liu$^2$, Jian Wu$^1$, Wenhui Duan$^1$\footnote{Author to whom
correspondence should be addressed. Email:
dwh@phys.tsinghua.edu.cn}, Bing-Lin Gu$^1$, and S. B. Zhang$^3$}
\affiliation{$^1$Center for Advanced Study and Department of
Physics, Tsinghua University, Beijing 100084, People's Republic of
China \\ $^2$College of Chemistry and Molecular Engineering,
Peking University, Beijing 100871, People's Republic of China \\
$^3$Department of Physics, Applied Physics, and Astronomy,
Rensselaer Polytechnic Institute, Troy, NY 12180, USA
}

\date{\today}
\begin{abstract}
First-principles calculations reveal half metallicity in zigzag
boron nitride (BN) nanoribbons (ZBNNRs). When the B edge, but not
the N edge, of the ZBNNR is passivated, despite being a pure
$sp$-electron system, the ribbon shows a giant spin splitting. The
electrons at the Fermi level are 100\% spin polarized with a
half-metal gap of 0.38 eV and its conductivity is dominated by
metallic single-spin states. The two states across at the Dirac
point have different molecular origins, which signals a switch of
carrier velocity. The ZBNNR should be a good potential candidate
for widegap spintronics.
\end{abstract}
\pacs{73.21.Hb, 73.22.-f, 71.15.Mb}

\maketitle

Half metallicity is at the forefront of spintronics
study\cite{Prinz,Wolf,Ziese}. Half metallicity occurs when one of
the electron spins shows insulating behavior while the other shows
metallic behavior\cite{Groot}. If one drives a current through such
a half metallic system, the current will be 100\% spin polarized.
Obviously, 100\% spin polarization could have many potential
spin-related applications\cite{Prinz,Wolf}. For sometime now, it is
understood that transition metal (TM)-containing systems such as
ferromagnetic manganese perovskite will show half
metallicity\cite{Park,Fang}. The TM systems, however, may not be
compatible with many of the matured technologies today that rely
heavily on main group semiconductors. Heavy TM elements also often
act as poison agents in biological systems. It is thus highly
desirable to develop non-TM half metallic materials, especially if
the half metallicity can be a byproduct of the existing electronic
materials. To this end, it is quite encouraging to see that
nanoscale zigzag graphene ribbons would show half metallicity under
high in-plane homogeneous electric field\cite{Son}, as graphene
ribbons could be an ideal conducting material for future
nanoelectronic applications\cite{Wakabayashi}. However, intrinsic half metallicity
without any external constraints is yet to be demonstrated and more
desirable in many practical applications.

In searching for intrinsic half metallicity in main group
semiconductors, we note that boron nitride (BN) nanoribbons may
hold high promises. BN nanotubes (BNNTs), hexegonal \emph{h}-BN,
and nanoribbons (BNNRs) are the III-V analogues of the widely
studied carbon nanotubes (CNTs), graphite, and graphene
nanoribbons (GNRs). Different from their carbon counterparts,
however, the BNNTs have a nearly constant band gap independent of
radius and chirality\cite{Blase}. The \emph{h}-BN is, on the other
hand, a wide gap semiconductor. Single layer \emph{h}-BN has been
successfully fabricated on the surfaces of metals
\cite{Nagashima}. The BNNRs are expected to be produced
straightforwardly by using single layer \emph{h}-BN as the
starting material, but should have very different physical
properties from those of \emph{h}-BN due to quantum size and
symmetry effects, and, as will be shown below, due to unexpected
edge effects. More importantly, the properties of the BNNRs may
also be qualitatively different from those of the GNRs, because of
the relatively large ionicity and significantly larger band gap of
the \emph{h}-BN.

In this paper, we predict intrinsic half metallicity in BNNRs with
zigzag edges (ZBNNRs) by first-principle calculations. The
half-metal energy gap for ribbons with passivated boron edge is as
high as 0.38 eV (about 15 times larger than $k_{\rm B}T$ at $T$ =
300 K). This does not require any applied external electric field,
in contrast to the graphene ribbons. Our analysis reveals that the
half metallicity is originated from an interesting interplay
between the nitrogen edge dangling bond state and the occupied
nitrogen lonepair state, which is absent in the graphene systems.
The crossing between the two states defines the Fermi level and
hence the degree of half metallicity. The intrinsically different
molecular orbital origins of the two states further suggest a
switch of the carrier velocity across the Dirac point that is
awaiting for experimental verification. The integration of half
metallicity (at the ribbon edge) with widegap semiconductivity (of
the ribbon backbone) also opens new application potentials whose
full extent is yet to be explored.

\begin{figure}
\includegraphics[width=0.8\textwidth]{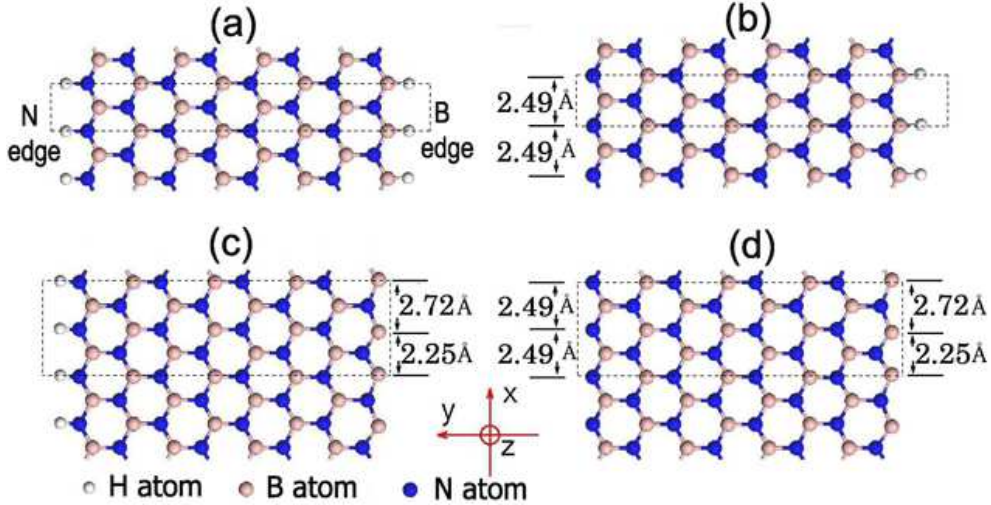}
\caption{(color online). Atomic geometries of the relaxed zigzag
boron nitride nanoribbons (ZBNNRs) with different edge treatments:
(a) 8-ZBNNR-2H, (b) 8-ZBNNR-HB, (c) 8-ZBNNR-HN, and (d) pristine
8-ZBNNR, where the number 8 stands for the width of the ribbon,
and the suffixes HB and HN stand for hydrogen termination of the
boron and nitrogen edges, respectively. The ribbons are infinitely
long along the $x$ direction. The distances between two adjacent
edge atoms are selectively shown in the figure. It should be noted
that in (c) and (d), the edge B atoms are dimerized, resulting in
two different edge B-B distances. The dashed rectangles denote the
unit cell of the systems.} \label{structure}
\end{figure}

The ZBNNRs we considered are schematically illustrated in Fig.
\ref{structure}, which are hexagonally bonded honeycomb ribbons
consisting of B and N atoms with zigzag terminated edges under
various passivations. In accordance with the previous convention
\cite{Fujita}, here the ZBNNRs are labeled by the number of
parallel zigzag chains, which defines the width of the ribbon. The
ZBNNR with $n$ B-N chains is thus named as $n$-ZBNNR. For ZBNNRs,
the outmost atoms at one edge (namely, the B-edge) are all B
atoms, whereas the outmost atoms at the other edge (namely, the
N-edge) are all N atoms. In terms of hydrogen passivation of the
edges, the ZBNNRs are further divided into four subgroups [Fig.
\ref{structure} (a)-(d)]: 1) both edges are passivated (ZBNNR-2H),
2) only the B-edge is passivated (ZBNNR-HB), 3) only the N-edge is
passivated (ZBNNR-HN), and 4) no edge is passivated (pristine
ZBNNR).

Our calculations were performed by using the density functional
theory \cite{note} within the local spin density approximation
(LSDA). In particular, we used the Vanderbilt planewave ultrasoft
pseudopotential \cite{Vanderbilt}, with a 450-eV cutoff energy,
and the Ceperly-Alder exchange-correlation potential \cite{CA}. We
adopted a supercell geometry for isolated BNNR sheet in which each
two adjacent sheets are separated by at least 11 \AA. For
$n$-ZBNNRs, the supercell contains $n$ [Fig. \ref{structure} (a)
and (b)] or $2n$ [Fig. \ref{structure} (c) and (d)] BN atoms
depending on whether dimerization of the edge atoms takes place or
not. Integration over the one dimensional Brillouin zone
($\Gamma$-$X$) has been carried out by using 51 and 31
Monkhorst-Pack $k$-points, respectively \cite{MP}, with the
equivalent k-point scheme. Full optimization of the atomic
structures including atomic positions and lattice parameters has
been carried out until the residual forces on atoms are less than
0.01 eV/\AA. We have also increased the size of the supercell to
make sure that it does not produce any discernible difference on
the results.

\begin{figure}[tbp]
\includegraphics[width=0.4\textwidth]{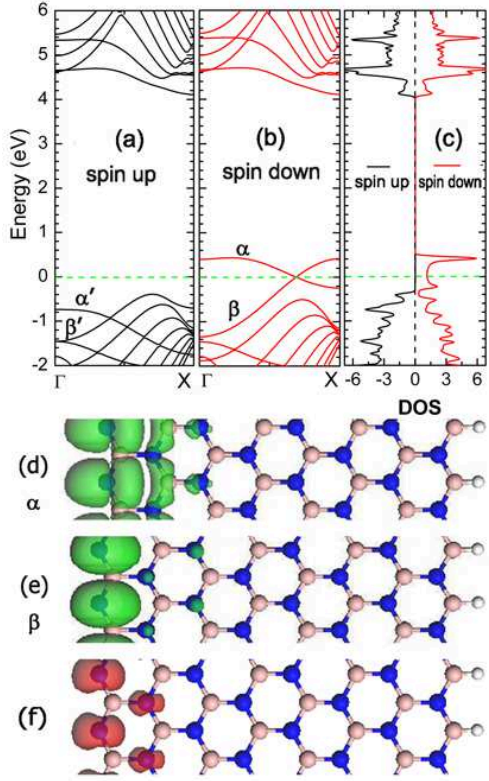}
\caption{(color online). Half metallic and ferromagnetic behaviors
of 8-ZBNNR-HB. (a) Spin-up and (b) spin-down energy bands, and (c)
the total density of states (DOS). Here, the energy zero (i.e.,
the dot-dashed green line) is at the Fermi level. $\Gamma$ and $X$
denote the center and the boundary of the first Brillouin zone.
(d) and (e) Partial charge density of the $\alpha$ and $\beta$
band at the $X$ point, respectively. The isosurface is 0.003
$e$/\AA$^3$. (f) Spatial distribution of the spin difference: red
for spin-up and yellow for spin down. The isosurface is 0.03
$\mu_B$/\AA$^3$. Due to the small amplitude of the spin-down
states, they are not visible from the plot. } \label{a8hb}
\end{figure}

The electronic and magnetic properties of the BN nanoribbons
depend critically on how the edges are passivated. Half
metallicity is observed when and only when the boron edge is
passivated [Fig. \ref{structure}(b)]. The band structures of the
8-ZBNNR-HB are depicted in Fig. \ref{a8hb}(a) and (b), which show
marked differences in the spin states: the spin-down electrons are
metallic with two bands ($\alpha$ and $\beta$) crossing each other
at the Fermi level, while the spin-up ones are insulating due to
the existence of a band gap as large as 4.5 eV. Thus, charge
transport are totally dominated by the spin-down electrons [see
the DOS in Fig. \ref{a8hb}(c)], and current flow in such a system
should be completely spin-polarized. The half-metal gap, defined
as the difference between the Fermi level and topmost occupied
spin-up band, is 0.38 eV. This value is comparable to that of
half-metallic graphene nanoribbon under high electric field
\cite{Son}, and is large enough for room-temperature operation. In
Fig. \ref{a8hb}(d) and (e), we plot the partial charge density of
the $\alpha$ and $\beta$ bands (at the characteristic $X$ point).
One can see that the $\alpha$ and $\beta$ bands are almost
entirely localized on the N edge. All the ZBNNRs-HB, regardless of
their width, have a similar band structure.

In order to understand the band crossing responsible for the
half-metallicity of ZBNNRs-HB, we have studied how passivation at
the N edge affects the electronic structure. With H passivation,
the $\alpha$ band shifts down considerably to reside inside the
valence band, while the $\beta$ band is very much unchanged and is
hence still above the top of the valence band. The projected
density of states (PDOS) of the two bands show that the $\alpha$
band is predominantly composed of the nitrogen $p_y$ and $s$
atomic orbitals, while the $\beta$ band is composed almost
completely of the nitrogen $p_z$ atomic orbitals. In other words,
the $\alpha$ band is a dangling bond state at the N edge, and is
hence strongly affected by the passivation. In contrast, the
$\beta$ band is the usual lonepair state of threefold coordinated
nitrogen, and is by-in-large unaffected by the passivation. The
wavefunctions of the $\alpha$ and $\beta$ bands have different
symmetries: being symmetric and anti-symmetric, respectively, with
respect to the basal plane of the BN sheet. These explain why the
$\alpha$ and $\beta$ bands in Fig. 2(b) cross each other without
having to create a band gap. If we bend the BN sheet or apply an
electric field along the $z$-direction to break such a mirror
symmetry, however, the $\alpha$ band will mix with the $\beta$
band, but the total energy of the system also increases. For example,
when the 8-ZBNNR-HB ribbon is bent to 90$^{\circ}$ around the $x$
axis, the total energy increases by 0.2 eV per unit cell. We have
checked the spin-up $\alpha^\prime$ and $\beta^\prime$ bands below
the Fermi level to find that they too have the same atomic characters
of the spin-down $\alpha$ and $\beta$ bands.

An essential factor underlying the observed half metallicity is
the splitting of the spin states. In fact, Fig. \ref{a8hb}(f) shows
that the ground state of the ZBNNRs-HB is \emph{ferromagnetic} with
a magnetic moment of 1 $\mu$B per edge N atom. The calculated magnetic
interaction is quite large: for instance, the ferromagnetic phase is
0.10 eV per edge N atom more stable than the anti-ferromagnetic phase,
and is also 0.17 eV per edge N atom more stable than the nonmagnetic
phase. Importantly, these energy differences are independent of the
ribbon width $n$ when $n\geq 5$. A relatively large distance (2.49 \AA)
between any two adjacent edge N atoms implies that no edge reconstruction
has taken place, and there is thus one dangling bond per edge N atom.
From such an analysis, we conclude that magnetism is a result of the
exchange interaction between dangling bond electrons. Similar magnetic
ordering in a dangling bond network was also observed on partially
hydrogenated Si(111) surfaces \cite{Okada}.

An obvious concern with the above discussion is the LDA band gap
error, which in principle could change the qualitative band
structure near the Fermi level in Fig. 2(b). To address such a
concern, we have calculated 3-ZBNNR-HB with the GW
approximation\cite{GW} by using the ABINIT code \cite{abinit}. The
results show that the system remains to be half-metallic with band
crossing at the Fermi level, despite significant quasiparticle
corrections to the LDA band structure, particularly in the
conduction bands\cite{supporting}.

In recent years, graphene has emerged as a new model system in
materials science and condensed matter physics due to its novel
physical properties \cite{GrapheneReview}. For example, graphene
is the first material where electron transport was found to be
governed by the relativistic Dirac equation: namely, energy
dispersion $E(k)$ with respect to wavevector $k$ is linear, so
that charge carriers mimic the relativistic quasiparticles with
zero rest mass (the so-called Dirac fermions) and travel with an
effective ``speed of the light'', $v = E/k$, on the order of $\sim
10^6$ m/s \cite{Novoselov}. Thus, it is important that our
ZBNNRs-HB also exhibits such an unusual massless Dirac-fermion
behavior. Due to the intrinsic difference between the $\alpha$ and
$\beta$ bands, however, the two bands at the crossing point will
have different slopes, corresponding to different $v$'s. This,
combined with the unique one-dimensional characteristics of the
edge states, suggests new physics that cannot exist in the
graphene systems with ``symmetric'' energy dispersion.

\begin{figure}[tbp]
\includegraphics[width=0.4\textwidth]{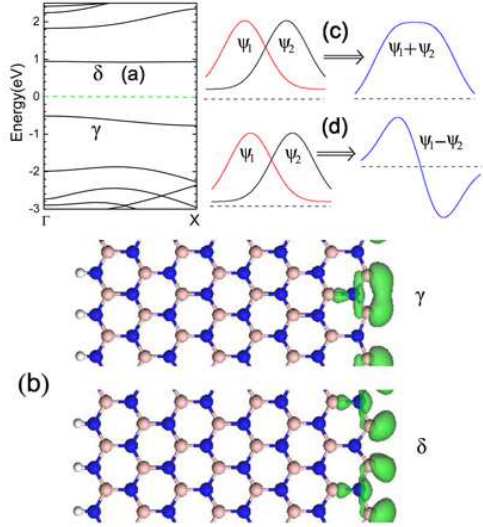}
\caption{(color online). Electronic properties and edge
dimerization of 8-ZBNNR-HN. (a) Energy band structure. The
$\gamma$ and $\delta$ bands are the highest valence and the lowest
conductance bands, respectively. Energy zero is at the nominal
Fermi level position (i.e., the green dashdotted line). (b)
Partial charge densities of the $\gamma$ and $\delta$ bands at the
$X$ point. The isosurface is 0.03 $e$/\AA$^3$. (c) and (d)
Schematic drawing of the formation of bonding and anti-bonding
orbitals from two adjacent atomic states.} \label{a8hn}
\end{figure}

When the N-edge, and only the N-edge, is passivated, the system,
ZBNNRs-HN, behaves qualitatively different. Figure 1c shows that
here a dimerization of the boron atoms at the bare B-edge takes
place spontaneously to lower the energy by 0.27 eV/dimer. The
system is no longer spin-polarized, but semiconducting with a gap
of 1.44 eV as shown in Fig. \ref{a8hn}(a). The B-B dimer has an
equilibrium bondlength of 2.25 {\AA}, which leaves the B-B
distance between two adjacent dimers to be 2.72 \AA. The
calculated partial charge densities of the highest occupied
$\gamma$ band and the lowest unoccupied $\delta$ band in Fig.
\ref{a8hn}(b) reveal that both states are localized at the bare B
edge. Bands $\gamma$ and $\delta$ exhibit the typical
characteristics of a bonding and anti-bonding orbital,
respectively. This suggests that the unpaired dangling bond states
of the boron atoms have rehybridized considerably. Note that edge
boron dimerization is a common phenomenon, which has been observed
in the simulation of the BNNT growth \cite{Growth} and for the
B-rich mouth of open zigzag BNNTs \cite{Shaogang}.

Dimerization is crucial for the disappearance of half metallicity.
If we prohibit the dimerization by artificially using a unit cell
that contains only $n$ B (N) atoms as those in Fig. 1(a) and (b),
the half metallicity and ferromagnetic behaviors will reappear.
For pristine ZBNNRs with no edge passivation, edge B atoms will
dimerize but edge N atoms will not, as shown in Fig. 1(d). This
system is also half metallic but with a negligible half-metal
energy gap of only 0.08 eV (see Supplementary Material for details
of the band structure\cite{supporting}). When both edges are
passivated, on the other hand, the system, e.g., ZBNNR-2H in Fig.
1(a), is nonspin-polarized and has a wide band gap. Our results
here qualitatively agree with the earlier work by Nakamura
\emph{et al.} \cite{Nakamura}.

To experimentally realize half metallicity in ZBNNRs, two key issues
must be addressed: (1) The choice of the substrate. Chemical activity
of the available substrates is often diverse, which should be utilized to
advance our course. For example, if one wishes to minimize the influence
of the substrate, an inert substrate should be used as was the case in
fabricating single layer \emph{h}-BN\cite{Nagashima}. (2) Selective
passivation. In the current study, hydrogen has been used as the passivant
for its simplicity. The drawback of using H is clear as the selectivity can
be rather poor. To optimize the selectivity, one might make good use of the
chemical difference between B and N. For example, a more electronegative
passivant such as F may be superior for boron passivation to yield the
desired half metallicity but may not work at all for nitrogen passivation.

In summary, first-principles study reveals half metallicity in ZBNNRs.
Specifically, boron edge passivated nanoribbons, ZBNNRs-HBs, is a half-metal
with a half-metal gap of 0.38 eV. Non-d ferromagnetism and completely spin-polarized
current transport may thus be possible. Nitrogen edge passivated nanoribbons, ZBNNR-HN
are, on the other hand, nonmagnetic and semiconducting. These unique properties,
especially the half-metallicity, make the BN nanoribbons attractive candidate for
nanoscale widegap spintronics such as spin-injection electrode, nano memory elements,
and nano transistors.

SBZ thanks Damien West for valuable discussions. This work was
supported by the National Natural Science Foundation of China
(Grant Nos. 10325415, 10674077, and 10774084) and the Ministry of
Science and Technology of China (Grant Nos. 2006CB605105 and
2006CB0L0601). SBZ was supported by the US DOE/BES and EERE under
the contract No. DE-AC36-99GO10337.

\end{document}